\documentclass[aps,pre,twocolumn,groupedaddress,showpacs,floatfix,superscriptaddress]{revtex4-1}
\usepackage[utf8]{inputenc}
\usepackage[T1]{fontenc}
\usepackage{graphicx}
\usepackage{amssymb}
\usepackage{nicefrac}
\usepackage{xcolor}
\usepackage{here}
\usepackage{tabularx}
\usepackage{bbm}

\usepackage{times}
\usepackage[colorinlistoftodos,prependcaption,textsize=small]{todonotes}

\usepackage{amsmath}
\DeclareMathOperator{\sign}{sign}

\usepackage{hyperref}

\hypersetup{
  colorlinks=true,
citecolor=black,
linkcolor=black,
urlcolor=black
  }

\newcommand{\bdt}[1]{\textcolor{black}{#1}}

\begin{document}
\preprint{AIP/123-QED}

\title{Underdamped, anomalous kinetics in double-well potentials}

\author{Karol Capa{\l}a}
\email{karol@th.if.uj.edu.pl} \affiliation{Institute of Theoretical Physics, and Mark Kac Center for Complex Systems
Research, Jagiellonian University, ul. St. {\L}ojasiewicza 11,
30--348 Krak\'ow, Poland}

\author{Bart{\l}omiej Dybiec}
\email{bartek@th.if.uj.edu.pl} \affiliation{Institute of Theoretical Physics, and Mark Kac Center for Complex Systems
Research, Jagiellonian University, ul. St. {\L}ojasiewicza 11,
30--348 Krak\'ow, Poland}

\date{\today}

\begin{abstract}
The noise driven motion in a bistable potential acts as the archetypal model of various physical phenomena.
Here, we contrast \bdt{properties of}  the overdamped \bdt{escape} dynamics with the full (underdamped) dynamics.
\bdt{In the weak noise limit}, for the overdamped particle driven by a non-equilibrium, $\alpha$-stable noise the ratio of forward and backward  transition rates depends only on the width of a potential barrier separating both minima.
Using analytical and numerical methods, we show that in the regime of full dynamics, contrary to the overdamped case, the ratio of transition rates depends both on widths and  heights of the potential barrier \bdt{separating minima of the double-well potential}.
The \bdt{derived} analytical formula for the ratio of transition rates is  corroborated by  extensive numerical simulations.
\end{abstract}


\maketitle

% \bd{Rownanie (9) sugeruje ze transition rate jest mniejsze o 1. Dlaczego nie mogłoby być większe od 1?}

% \bd{Czy wiesz co dzieje sie dla $\alpha=2$? Jest jakiś wzór?}

%%%%%%%%%%%%%%%%%%%%%%%%%%%%%%%%%%%%%%%%%%%%%%%%%%%%%%%%%%%%%%%%%%%%%%%%%%%%%%%%%%%%%%
%%%%%%%%%%%%%%%%%%%%%%%%%%%%%%%%%%%%%%%%%%%%%%%%%%%%%%%%%%%%%%%%%%%%%%%%%%%%%%%%%%%%%%
\section{Introduction}

%\bdt{Gdzieś trzeba zacytować: \cite{Xu2020,wang2020}}

A noise induced escape of a particle is one of archetypal problems in stochastic dynamics.
It underlines various noise driven effects.
Among others, it was  studied by H.~A.~Kramers in the case of the Gaussian white noise (GWN) in overdamped (large viscosity) and underdamped (small viscosity) regimes \cite{kramers1940}.
In these cases, the ``velocity of chemical reactions''  (reaction rate) depends only on the height of the barrier separating reactants.
Moreover, in the overdamped regime, the obtained formula for the reaction rate can be interpreted as the Arrhenius equation \cite{arrhenius1889}.
Therefore, the stochastic motion in the double-well potential can be used as an effective model of chemical reactions.
Since then, the noise induced escape of a particle was intensively studied in the overdamped \cite{hanggi1990,mel1991kramers} and underdamped \cite{melnikov1986,mel1991kramers,Shneidman1997,Barcilon1996} regimes as well as in quantum setups  \cite{rips1986,rips1990quantum,topaler1994quantum}.

The Gaussian white noise is a very special representative of the more general family of $\alpha$-stable white noises.
Except the Gaussian white noise, $\alpha$-stable noises have the so-called ``heavy tails'', i.e., they allow for occurrence of extreme events with a significantly larger probability than the Gaussian distribution.
For instance, noise induced displacements under L\'evy noises with $0< \alpha < 2$ follow the power-law distribution with the exponent $-(\alpha+1)$.
Consequently, only fractional moments of order $\nu$ which is smaller than $\alpha$ exist \cite{samorodnitsky1994,janicki1996}, i.e. $\langle |x|^\nu \rangle  < \infty $.
\bdt{Power-law}, heavy-tails of $\alpha$-stable densities are responsible not only for divergence of moments, but also for discontinuity of paths of processes driven by L\'evy noises \cite{janicki1994b}.
In particular, in the overdamped and underdamped regime, position or velocity, respectively, is discontinuous.
\bdt{Finiteness of higher order moments can be reintroduced by the so called truncated (tempered)  L\'evy flights \cite{sokolov2004,rosinski2007,mantegna1994b,shlesinger1995b,koponen1995,nakao2000,kuchler2013}.}

% In particular, in the overdamped regime, the position is discontinuous, while in the underdamped regime the velocity is discontinuous.

Heavy-tailed, L\'evy type fluctuations, similarly to the equilibrium, thermal GWN noise, leads to many surprising noise-induced phenomena like ratcheting effect \cite{magnasco1993,reimann2002,li2017transports}, stochastic resonance \cite{gammaitoni2009} or resonant activation \cite{doering1992}.
% Moreover, non-negligible probability of extreme events produces multimodal stationary states in single-well potentials steeper than parabolic \cite{chechkin2002,chechkin2003,capala2019multimodal,capala2019underdamped}.
% \karol{Rozważyć, czy zachowujemy powyższe zdanie.}
Non-Gaussian, heavy-tailed fluctuations have been observed in plenitude of experimental setups ranging from
disordered media \cite{bouchaud1990}, biological systems \cite{bouchaud1991}, rotating flows \cite{solomon1993}, optical systems and materials \cite{barthelemy2008,mercadier2009levyflights}, physiological applications \cite{cabrera2004},
financial time series \cite{laherrere1998,mantegna2000,lera2018gross},  dispersal patterns of humans and animals \cite{brockmann2006,sims2008}, laser cooling \cite{barkai2014} to
 gaze dynamics \cite{amor2016} and search strategies \cite{shlesinger1986,reynolds2009}.
\bdt{
They are studied both experimentally \cite{solomon1993,solomon1994,amor2016} and theoretically \cite{metzler2000,barkai2001,chechkin2006,jespersen1999,klages2008,dubkov2008} including the problem of fluctuation-dissipation relations in non-equilibrium systems \cite{touchette2007,touchette2009,chechkin2009,dybiec2012,kusmierz2014}.
Consequently, despite some nonphysical features, due to well-known mathematical properties, e.g., self similarity, infinite divisibility and generalized central limit theorem, $\alpha$-stable noises are widely applied in various models displaying anomalous fluctuations or describing anomalous diffusion.
}
\bdt{One may also consider their more physical counterparts, namely L\'evy walks \cite{zaburdaev2015levy}, for which ``long jumps'' are performed with finite velocity.
Despite this difference, such systems can still exhibits some similar phenomena to L\'evy flight \cite{Xu2020,wang2020} }

One might expect that $\alpha$-stable noise can significantly change properties of escape kinetics in overdamped systems.
Indeed, contrary to the Gaussian white noise driving, for which the rate of reaction rates depends only on the depth of the potential well \cite{kramers1940}, under $\alpha$-stable noise the ratio of transition rates \bdt{is sensitive to}  the width of the potential barrier \cite{ditlevsen1999,imkeller2006,imkeller2006b,bier2018}.
In the weak noise limit, i.e., when the noise intensity tends to $0$, the dependence of the ratio of transition rates solely on the width of the potential barrier can be demonstrated \cite{imkeller2006,imkeller2006b}.
This relation holds also for finite noise intensity, as long as noise intensity is much smaller that the depth of the potential well \cite{chechkin2005}, however the combined action of the L\'evy noise and the Gaussian noise might reintroduce the sensitivity of the ratio of transition rates to the barrier height \cite{capala2020athermal}.

In the regime of full dynamics, a particle is characterized both by the velocity and the position.
Depending on the noise type, the velocity can be discontinuous, e.g., for L\'evy noises with $\alpha<2$.
At the same time, the position is continuous, which might change properties of the same models in comparison to their overdamped counterparts.
In this manuscript, we extend the discussion on the underdamped kinetics driven by L\'evy noises in double-well potentials.
In the next section (Sec.~\ref{sec:model} Model) we derive the relation between transition rates in the weak noise limit \bdt{given by Eq.~(\ref{eq:transitionRate}), which is the main result of this manuscript.}
In the Sec.~\ref{sec:results} (Results) we present extensive comparisons between the derived approximate formula obtained in Sec.~\ref{sec:model} and results of numerical simulations.
The manuscript is closed with Summary and Conclusions (Sec.~\ref{sec:summary}).

%%%%%%%%%%%%%%%%%%%%%%%%%%%%%%%%%%%%%%%%%%%%%%%%%%%%%%%%%%%%%%%%%%%%%%%%%%%%%%%%%%%%%%
%%%%%%%%%%%%%%%%%%%%%%%%%%%%%%%%%%%%%%%%%%%%%%%%%%%%%%%%%%%%%%%%%%%%%%%%%%%%%%%%%%%%%%
\section{Model \label{sec:model}}

The Langevin equation \cite{gardiner1983} provides description of a particle motion in a noisy environment.
In the underdamped regime, the Langevin equation takes the following form
\begin{equation}
    m \ddot{x}(t) = -\gamma \dot{x}(t)  -V'(x) + \sigma \zeta(t),
    \label{eq:langevin}
\end{equation}
where $-V'(x)$ is the deterministic force acting on a particle, while $\zeta(t)$ stands for the noise (random force), which approximates interactions of the test particle with its environment.
\bdt{The scale parameter $\sigma$ ($\sigma>0$) controls the strength of fluctuations.}
\bdt{In Eq.~(\ref{eq:langevin}), $x$ has the dimension of length, $t$ of time, $[V(x)]=[\mbox{energy}]$.
Remaining parameters have following units:  
$[\gamma]=\mbox{mass}/\mbox{time}$,
$[\sigma]= \mbox{length} \times \mbox{mass} / (\mbox{time})^{1+\frac{1}{\alpha}}$
and
$[\zeta]=\mbox{time}^{\frac{1}{\alpha}-1}$.
}
We assume that the noise $\zeta(t)$ is of white, $\alpha$-stable, L\'evy type, i.e., it generalizes the Gaussian white noise \cite{samorodnitsky1994,janicki1996}.
\bdt{Therefore, for $\alpha<2$, $\sigma$ and $\gamma$, contrary to $\alpha=2$, are two independent parameters.}
Moreover, we restrict ourselves to symmetric $\alpha$-stable noise only, which
is the formal time derivative of the symmetric $\alpha$-stable motion ${L}(t)$, see~\cite{janicki1994b}, whose characteristic function is given by
\begin{equation}
 \phi(k)=\left\langle \exp[ik {L}(t)] \right\rangle=\exp\left[ -  t  |k|^\alpha \right].
 \label{eq:fcharakt}
\end{equation}
The stability index  $\alpha$ ($0<\alpha \leqslant 2$) controls the noise asymptotics.
For $\alpha=2$, the $\alpha$-stable noise transforms into the standard Gaussian white noise \cite{samorodnitsky1994,janicki1996}.
Increments of the symmetric $\alpha$-stable motion $ {L}(t)$, i.e., $\Delta  {L} =  {L}(t+\Delta t)- {L}(t)$, \bdt{are independent, identically, distributed} according to a symmetric $\alpha$-stable density with the characteristic function given by Eq.~(\ref{eq:fcharakt}) with $t$ replaced by $\Delta t$.
Importantly, symmetric $\alpha$-stable densities are unimodal probability densities which for $\alpha<2$ exhibit a power-law asymptotics with tails decaying as $|\zeta |^{-(\alpha+1)}$, see \cite{samorodnitsky1994,janicki1996}.
Consequently, for $\alpha<2$, all moments of order greater than $\alpha$, e.g., variance, diverge.

\begin{figure}[h!]
\centering
\includegraphics[width=0.99\columnwidth]{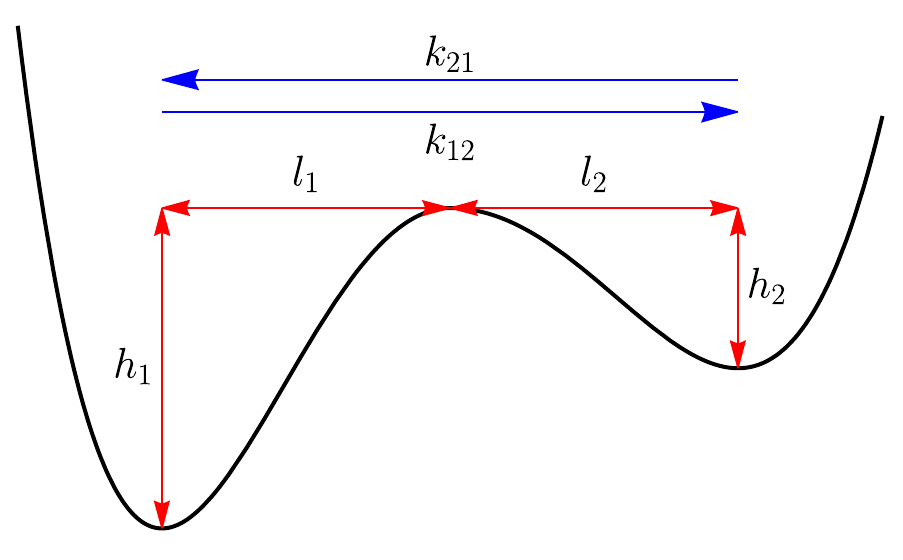}
\caption{Schematic sketch of the potential, see  Eq.~(\ref{eq:potential}), used in numerical studies of \bdt{noise induced} escape kinetics.
}
\label{fig:potencial}
\end{figure}

Equation~(\ref{eq:langevin}) can be rewritten as the set of two first order equations
\begin{equation}
\left\{
\begin{array}{l}
m\dot{v}(t)=-\gamma v(t) - V'(x) + \sigma \zeta(t) \\
    \dot{x}(t)=v(t)
\end{array}
\right..
\label{eq:fullLangevin}
\end{equation}
The deterministic force $-V'(x)$ is produced by the fixed, double-well, potential $V(x)$, with two minima located at $x_1$ and $x_2$ ($V(x_1)=E_1$ and $V(x_2)=E_2$) and a single local maximum at $x_b$ ($x_1<x_b<x_2$ and $V(x_b)=E_b$), \bdt{see Fig.~\ref{fig:potencial}}.
\bdt{Without the loss of generality it can be assumed that $x_b=0$ and $V(x_b)=0$.}
% \karol{Niebezpieczeństwo: potencjał na Fig.~\ref{fig:potencial} ma $h_i=\Delta E_i$ a nie $E_i$}
Furthermore, we assume that both, the potential barrier separating potential minima and outer (large $|x|$) parts of the potential are steep enough to assure that the particle position is limited to the neighborhood of potential minima.

Using Eq.~(\ref{eq:fullLangevin}), we study the problem of noise induced escape over the \bdt{static} potential barrier, with the special attention to the weak noise limit.
Under the weak noise approximation, the L\'evy noise can be \bdt{effectively} decomposed into the Wiener part \bdt{(small, bounded jumps)} and the compound Poisson process \bdt{(spikes)} \cite{imkeller2006,imkeller2006b}.
\bdt{
More precisely, the L\'evy--Khintchine formula \cite{zolotarev1986} shows that a L\'evy process $L(t)$ is built by three independent components: a linear drift (deterministic motion), a Brownian motion and a L\'evy jump process.
Furthermore, in \cite{imkeller2006,imkeller2006b}, it has been shown that for the small $\sigma$ it is possible to introduce a threshold $\delta(\sigma)$ such that all subthreshold pulses are considered as background, while  suprathreshold  pulses build spikes.
The small jumps part makes infinitely many jumps on any time interval of positive length, but the absolute  value of these jumps is bounded.
In \cite{imkeller2006,imkeller2006b}, it has been proved that, for appropriately chosen $\delta(\sigma)$,  the variance of the background (small jumps) part vanishes in the limit of $\sigma\to 0$.
Consequently, between the two subsequent large spikes, the particle is subjected only to a background noise, which, for small $\sigma$, is so weak that the motion of a particle is almost deterministic.
Moreover, time lags between the two subsequent spikes are so large that the particle practically reaches the bottom of the potential well (underdamped dynamics) or velocity drops almost to zero (full dynamics).
Therefore, in the weak noise limit, the only scenario capable of inducing the escape is when a strong enough spike ``kicks'' the particle.
Importantly, the weak noise regime is recorded already for finite, although small, values of the scale parameter $\sigma$.
The exact value of the $\sigma$ for which agreement with predictions corresponding to $\sigma\to 0$ is recorded depends on the setup under study.
The detailed discussion of the decomposition procedure can be found in Refs.~\cite[Sec.~2]{imkeller2006}, \cite[Sec.~3]{imkeller2006b} or \cite[Sec.~3.1]{pavlyukevich2010}.
}
\bdt{In overall}, the  bounded jump component part is responsible for short displacements, while the Poisson part controls long jumps.
For the L\'evy noise, characterized by the stability index $\alpha$ ($0<\alpha<2$), the probability of recording an event $\xi$ larger than $\zeta$ is given by
\begin{equation}
    P(\xi>\zeta)\sim \zeta ^{-\alpha}.
    \label{eq:tails}
\end{equation}

% In further considerations, we will  disregard the Gaussian component, as properties of escape kinetics in the weak noise limit are mainly determined by tails of the jump length distribution.
% \bdt{This can be justified by the fact that we are considering the weak noise ($\sigma\to 0$) for which the bounded part itself is very unlikely to induce a transition over the potential barrier.
% The transition over the potential barrier is produced by a single strong pulse in the velocity ruled  by a compound Poisson part.
% For a detailed discussion see Refs.~\cite[Sec.~2]{imkeller2006}, \cite[Sec.~3]{imkeller2006b} or \cite[Sec.~3.1]{pavlyukevich2010} }

\bdt{In the weak noise limit}, the protocol of escaping over the potential barrier is based on a single long ``jump'' in the velocity, which in a single, strong ``kick'', gives the particle kinetic energy sufficient to overpass the potential barrier deterministically.
More precisely, we assume that initially a particle has velocity $v_0$, and it is located in the $i$th minimum of the potential.
From this point, it moves deterministically to the top of the potential barrier.
During the motion to the top of the barrier, it loses some of its energy due to the friction.
% Simultaneously, the kinetic energy is transformed into the potential one.
\bdt{Moreover, it is perturbed by the small jumps component, which typically is weak enough not to suppress the transition over the potential barrier.}
If we disregard the friction, the minimal velocity, which is sufficient to produce the transition from the $i$th minimum to the barrier top, reads
\begin{equation}
    \frac{mv^2}{2}\geqslant E_b-E_i=\Delta E_i.
    \label{eq:crossingCondition}
\end{equation}
During the motion the energy is dissipated by friction, therefore, the minimal initial velocity $v_0$ needs to be larger
\begin{equation}
    v_0 = v + \frac{\gamma}{m} \int_{t_0}^{t_0+\delta t} v(t) dt,
\end{equation}
where $\delta t$ ($\delta t \gg 0$) is the time necessary to reach the top of the potential barrier.
The integration over time gives the distance between the initial position $x_i$ and the potential barrier $x_b$, i.e., $l_i$.
Consequently,  the initial velocity reads
\begin{equation}
    v_0=v+\frac{\gamma}{m} l_i.
    \label{eq:vFinal}
\end{equation}
Combining Eqs.~(\ref{eq:crossingCondition}) and (\ref{eq:vFinal}), we get the following estimate for the minimal initial velocity $v_0$
\begin{equation}
    v_0=\sqrt{\frac{2\Delta E_i}{m}} + \frac{\gamma}{m} l_i.
    \label{eq:v0}
\end{equation}

Equation~(\ref{eq:langevin}) describes the full (underdamped) dynamics in the regime of linear damping.
For a free particle under linear friction, the velocity is distributed according to the $\alpha$-stable density with the same stability index $\alpha$ as the noise, \bdt{see Refs.~\cite{chechkin2002,chechkin2003,dybiec2007d} and Appendix~\ref{sec:velocity}}.
The deterministic force $-V'(x)$, see the first line of Eq.~(\ref{eq:fullLangevin}), affects the shape of the stationary velocity distribution.
For the weak noise, i.e., small $\sigma$, the majority of particles are localized in the vicinity of potential minima, where the deterministic force is small and can be neglected.
Consequently, in the weak noise limit, we can assume that the velocity is distributed according to the $\alpha$-stable density, while for the larger $\sigma$ it can be approximated by the $\alpha$-stable density.
\bdt{Please note, that such a situation corresponds to the diverging mean energy $\langle E \rangle$, because $\alpha$-stable densities with $\alpha<2$ are characterized by the diverging variance. The divergence of mean energy does not affect our considerations, because we calculate the probability of recording a minimal instantaneous energy.
The condition on the minimal instantaneous energy can be transformed into the equivalent condition on the instantaneous velocity, see Eq.~(\ref{eq:transitionProbability}), which is easier to utilize, due to known asymptotic behavior of $\alpha$-stable densities, see Eq.~(\ref{eq:tails}) and~(\ref{eq:asymptoti}).}
As the first approximation, we assume that the large initial velocity is directed towards the potential barrier.
If the potential barrier is narrow, and outer parts of the potential are steep, the particle is unlikely to explore positions placed beyond minima, i.e., $|x| \gg |x_i|$.
Consequently, the large velocity is most likely to be directed towards the potential barrier.
As it will be shown later, the transition initiated by the abrupt velocity towards the potential barrier is the most probable and the approximation based on this assumption, see Eq.~(\ref{eq:transitionRate}), works very well.
If the velocity is not large enough, the particle could be reversed prior to reaching the top of the potential barrier.
On the one hand, transitions over the potential barrier are produced by extreme velocities, which are ruled by the tail of the velocity distribution.
On the other hand, a particle during its motion to the barrier top is subject to damping and to continuous small perturbations, controlled by the central part of the $\alpha$-stable density, i.e., the Gaussian like part.
Employing Eq.~(\ref{eq:tails}), we find the probability that the velocity larger than the minimal value $v_0$ is recorded
\begin{equation}
    P(v>v_0) \sim \left( \sqrt{\frac{2\Delta E_i}{m}} + \frac{\gamma}{m} l_i \right) ^{-\alpha}.
    \label{eq:transitionProbability}
\end{equation}
If the initial position of the particle is in the $i$th minimum, i.e., $x(t_0)=x_i$, the probability given by Eq.~(\ref{eq:transitionProbability}) is equal to the transition rate $k_{ij}$.
Therefore the ratio, $\kappa$, of forward, $k_{12}$, and backward, $k_{21}$, transition reads
\begin{equation}
    \kappa=\frac{k_{12}}{k_{21}}=\left(\frac{\sqrt{2\Delta E_2} + \gamma \sqrt{m} l_2}{\sqrt{2\Delta E_1} + \gamma \sqrt{m} l_1}\right)^{\alpha}.
    \label{eq:transitionRate}
\end{equation}
The derivation of Eq.~(\ref{eq:transitionRate}), assumes that the particle is wandering around a minimum of the potential and waiting for the extreme velocity larger than $v_0$, see Eq.~(\ref{eq:v0}).
If the particle velocity is larger than $v_0$, it can overpass the potential barrier practically in the deterministic manner.
For $\gamma \to \infty$, Eq.~(\ref{eq:transitionRate}) reduces to the well-known overdamped limit, where the ratio of transition rates depends only on the ratio of distances between potential minima and the barrier top \cite{ditlevsen1999, imkeller2006, imkeller2006b}, i.e.,
\begin{equation}
     \kappa=\frac{k_{12}}{k_{21}}=  \left(\frac{l_2}{l_1}\right)^{\alpha}.
     \label{eq:overdampedRatio}
\end{equation}

\bdt{For weak enough noise (small $\sigma$)}, transition rates and their ratio can be calculated using the relationship with the mean first passage times (MFPT), see \cite{hanggi1990}.
For the particle starting in the left minimum $x_1$ of the potential the MFPT, $T_{12}$, is defined as
\begin{eqnarray}
    T_{12}  =  \langle \tau \rangle =
     \langle \min\{\tau : x(0)=-l_1 \;\land\; x(\tau) \geqslant 0  \} \rangle.
    \label{eq:mfpt-def-forward}
\end{eqnarray}
Therefore, the forward transition rate, $k_{12}$, is given by
\begin{equation}
k_{12}=\frac{1}{T_{12}}.  
\end{equation}
Definitions of the MFPT from the right potential well, $T_{21}$, and the backward transition rate, $k_{21}$, are analogous \bdt{to the definition of $T_{12}$ and $k_{12}$.}
Finally, from numerically estimated MFPTs the ratio of transition rates can be calculated
\begin{equation}
  \kappa=\frac{k_{12}}{k_{21}} = \frac{T_{21}}{T_{12}}  .
\end{equation}
The mean first passage times $T_{12}$ and $T_{21}$ can be obtained using \bdt{numerical} simulations of the Langevin equation, which can be rewritten in the discretized form
\begin{equation}
    \left\{
\begin{array}{l}
v_{i+1}= v_i -\left[(\gamma v_i + V'(x_i))\Delta t + \sigma \left(\Delta t\right)^{1/\alpha}\zeta_i \right]/m \\
x_{i+1}=x_i+v_{i+1} \Delta t
\end{array}
\right.,
    \label{eq:lanngevinDiscrete}
\end{equation}
\bdt{where $\zeta_i$ is the sequence of independent, identically distributed $\alpha$-stable random variables and $\Delta t$ is the integration times step, which is significantly smaller than the transition time, i.e., $\Delta t \ll \delta t$.}
The velocity part, containing the $\alpha$-stable noise, is approximated using the Euler-Maruyama scheme \cite{janicki1994,janicki1996}, while the spatial part is constructed trajectory-wise.
In order to estimate the required MFPT $T_{12}$ ($T_{21}$) trajectories $x(t)$ are generated using the approximation (\ref{eq:lanngevinDiscrete}), with the initial condition $x(0)=-l_1$ ($x(0)=l_2$) \bdt{and $v(0)=0$}, as long as $x(t) < x_b$ ($x(t) > x_b$).
From the ensemble of first passage times, the mean first passage times and their ratios are calculated.
Within computer simulations, it is assumed that the particle mass is set to $m=1$.

The approximation given by Eq.~(\ref{eq:transitionRate}) suggests that the ratio of escape rates depends both on depths of potential wells and distances between minima and the maximum of the potential.
Therefore, we use such a potential which allows easy control of its depths and distances between minima and the maximum
\begin{equation}
  V(x) =  \left\{
\begin{array}{l}
 4 h_1 \left[\frac{x^4}{4 l_1^4}-\frac{x^2}{2 l_1^2}\right] \quad x<0 \\
 \\
 4 h_2 \left[\frac{x^4}{4 l_2^4}-\frac{x^2}{2 l_2^2}\right] \quad x \geqslant 0
\end{array}
\right..
\label{eq:potential}
\end{equation}
\bdt{Parameters} $h_1$ and $h_2$ controls depths of the left and right minimum respectively, while $l_1$ and $l_2$ represent distances between the potential maximum and the corresponding minimum.
\bdt{The top of the potential barrier is located at $x_b=0$.}
The potential given by Eq.~(\ref{eq:potential}) is schematically depicted in Fig.~\ref{fig:potencial}.

Numerical results were obtained by use of the discretized version of the Langevin equation, see Eq.~(\ref{eq:lanngevinDiscrete}).
Simulations were performed mainly with the integration time step $\Delta t=10^{-3}$, which is significantly smaller than the transition time $\delta t$.
Nevertheless, some of them were repeated with the smaller integration time step, i.e., $\Delta t=10^{-4}$.
\bdt{Such an integration time step was sufficient to ensure stability of the Euler-Maruyama method.}
Final results were averaged over $N=10^5 - 10^6$ realizations.
For simplicity, we have assumed $m=1$ and $\gamma=1$ (except situations when it is varied).
Remaining parameters: $l_1$, $l_2$, $h_1$, $h_2$ and $\sigma$ varied among simulations.
Their exact values are provided within the text and figures' captions.

%%%%%%%%%%%%%%%%%%%%%%%%%%%%%%%%%%%%%%%%%%%%%%%%%%%%%%%%%%%%%%%%%%%%%%%%%%%%%%%%%%%%%%
%%%%%%%%%%%%%%%%%%%%%%%%%%%%%%%%%%%%%%%%%%%%%%%%%%%%%%%%%%%%%%%%%%%%%%%%%%%%%%%%%%%%%%

% \clearpage
\section{Results\label{sec:results}}

\begin{figure}[h!]
\centering
\includegraphics[width=0.99\columnwidth]{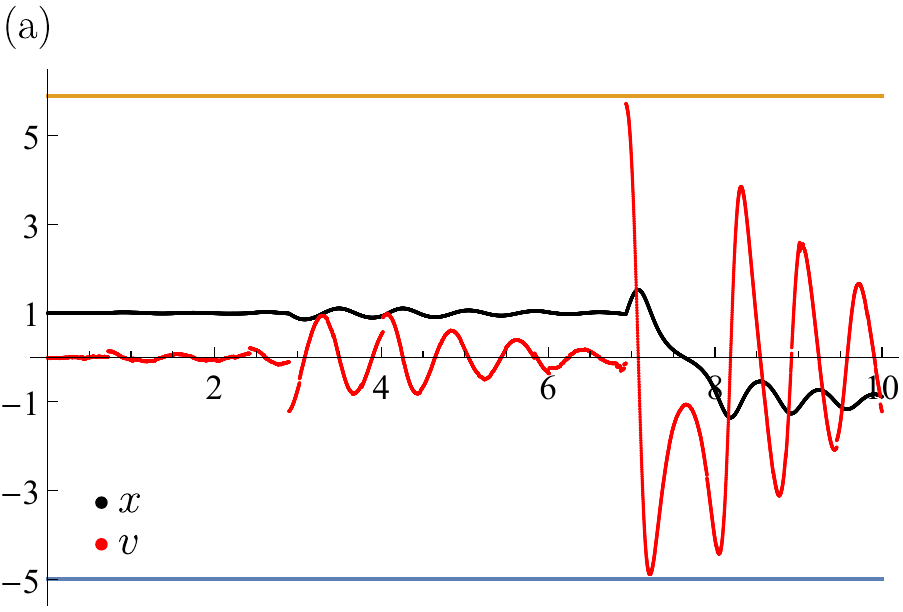} \\
\includegraphics[width=0.99\columnwidth]{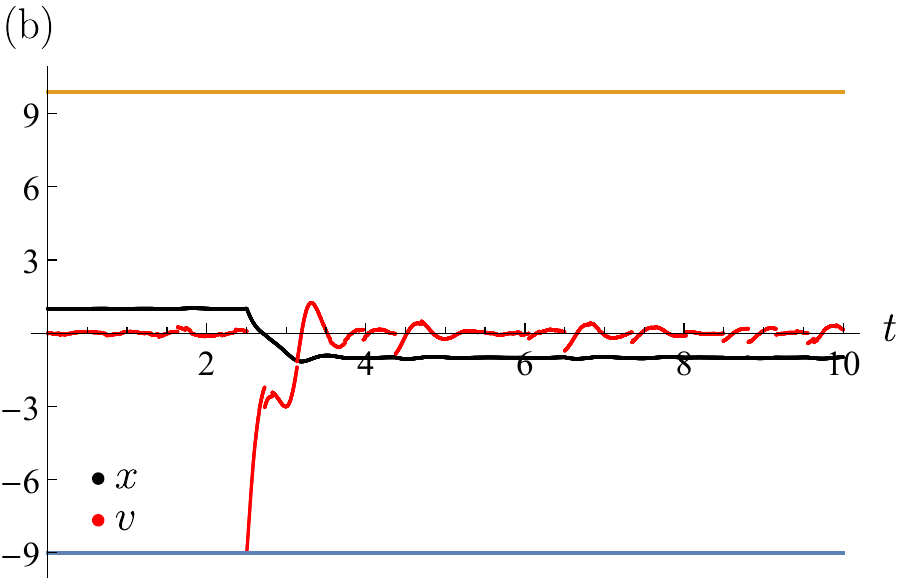}
\caption{Sample trajectories of the particle moving in the potential (\ref{eq:potential}) with $l_1=l_2=1$, $h_1=12$ and $h_2=8$. The stability index $\alpha$ is equal to $\alpha=1$ and the damping coefficient $\gamma$ is set to $\gamma=1$ (top panel --- (a)) and $\gamma=5$ (bottom panel --- (b)).
Horizontal lines show minimal velocities for forward (orange) and backward (blue) transitions which are given by Eq.~(\ref{eq:v0}).
\bdt{More details in the text.}
}
\label{fig:trajectories}
\end{figure}

We start our studies with the inspection of trajectories of the process generated by Eq.~(\ref{eq:langevin}) under Cauchy ($\alpha=1$) noise, see Fig.~\ref{fig:trajectories}.
The top panel shows results for $\gamma=1$, while in the bottom panel the damping is set to $\gamma=5$.
Since the motion is perturbed by the $\alpha$-stable noise, the velocity $v(t)$ is discontinuous, while  the position $x(t)$, $x(t)=\int v(t)dt$, is continuous.
First of all, with the increasing damping, the particle motion becomes more restricted, i.e., the particle is most likely to be found in the vicinity of one of the potential wells
\bdt{because position fluctuates less.}
At the same time, the particle loses its velocity and energy faster,
\bdt{what is manifested by faster decay and lower amplitude of velocity oscillations in the bottom panel.}
Inspection of trajectories confirms that, in order to overpass the potential barrier, the instantaneous velocity needs to be large enough and, interestingly, it can be directed both towards the top of the potential barrier (bottom panel) or outwards (top panel).
Horizontal lines in Fig.~\ref{fig:trajectories} depict minimal values of velocities \bdt{towards the potential barrier}, see Eq.~(\ref{eq:v0}), which are sufficient to induce a transition over the potential barrier.
Due to the potential asymmetry, minimal forward (\bdt{from the left to the right}) and backward (\bdt{from the right to the left}) velocities are different.
Moreover, because of the damping, the minimal velocity in the direction of the boundary is smaller than the minimal velocity in the opposite direction.

The top panel of Fig.~\ref{fig:trajectories} shows the situation when the initial large velocity is pointing in the opposite direction than the potential barrier.
\bdt{After a strong noise pulse at $t\approx 7$}, a particle initially moves to the right.
It gets to the reversal point, in which the velocity drops to zero and the motion is reversed.
The particle returns to the right minimum of the potential, where it has the negative velocity equal to the minimal backward velocity.
Consequently, it continues its motion towards the potential barrier, which is successfully overpassed.
% In the vicinity of the top of the potential barrier, the absolute value of the velocity is minimal.
After passing the potential barrier, due to the deterministic force, the particle accelerates.
In the bottom panel of Fig.~\ref{fig:trajectories} the initial velocity \bdt{after a strong pulse at $t\approx 2.5$} is equal to the minimal backward velocity and it is directed towards the potential barrier.
Consequently, the particle can successfully pass from the right to the left minimum of the potential.
Moreover, during the sliding from the barrier top to the left minimum of the potential, the velocity is perturbed \bdt{a couple of times, e.g., at $t\approx 2.72$, $t \approx 2.82$  and $t\approx 3.1$ discontinuities in $v(t)$ are visible.}
Fig.~\ref{fig:trajectories} clearly confirms that the assumption of the ``single-jump'' escape is fully legitimate.

\begin{figure}[h!]
\centering
\includegraphics[width=0.95\columnwidth]{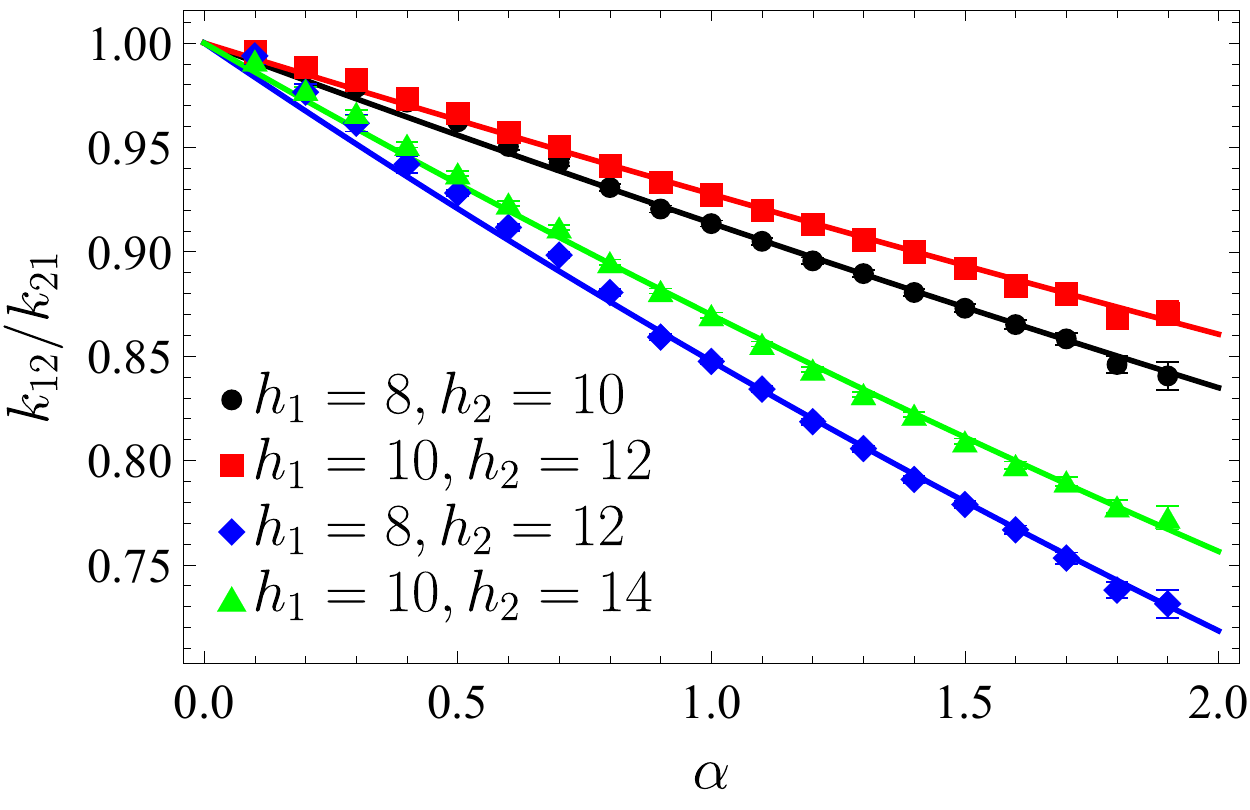}
\caption{Ratio $\kappa(\alpha)$ of transition rates from minima of the potential (\ref{eq:potential}) to the barrier top as a function of the stability index $\alpha$.
Various points correspond to numerical results for different depths of potential wells, i.e., different values of  $h_1$ and $h_2$, while lines plot the scaling given by Eq.~(\ref{eq:transitionRate}).
Simulation parameters $l_1=1$,  and $l_2=1$, $\gamma=1$ and $\sigma=0.2$.
}
\label{fig:rate}
\end{figure}

\bdt{Figure~\ref{fig:rate} compares numerically calculated ratio of transition rates (points) with predictions of Eq.~(\ref{eq:transitionRate}) (lines) as a function of the stability index $\alpha$.}
It corresponds to fixed distances between minima and the maximum of the potential ($l_1=1=l_2=1$) and various potential depths ($h_1$ and $h_2$).
Parameters $h_1$ and $h_2$, characterizing depths of potential wells, were chosen in such a way that both are of the same order, and \bdt{their values are} significantly larger than the scale parameter $\sigma=0.2$, i.e., $h_1 \gg 0.2$ and $h_2 \gg 0.2$.
Such a choice of parameters ensures that weak noise approximation can be employed.
Fig.~\ref{fig:rate} clearly shows that the ratio of transition rates depends on depths of both potential wells.
For $\alpha>1$, there is a perfect agreement between results of simulations and the formula (\ref{eq:transitionRate}).
For small values of the stability index $\alpha$ ($\alpha<1$), there are some discrepancies.
More precisely, the numerically estimated ratio of transition rates is slightly larger than the expected scaling given by Eq.~(\ref{eq:transitionRate}).

\begin{figure}[h!]
\centering
\includegraphics[width=0.95\columnwidth]{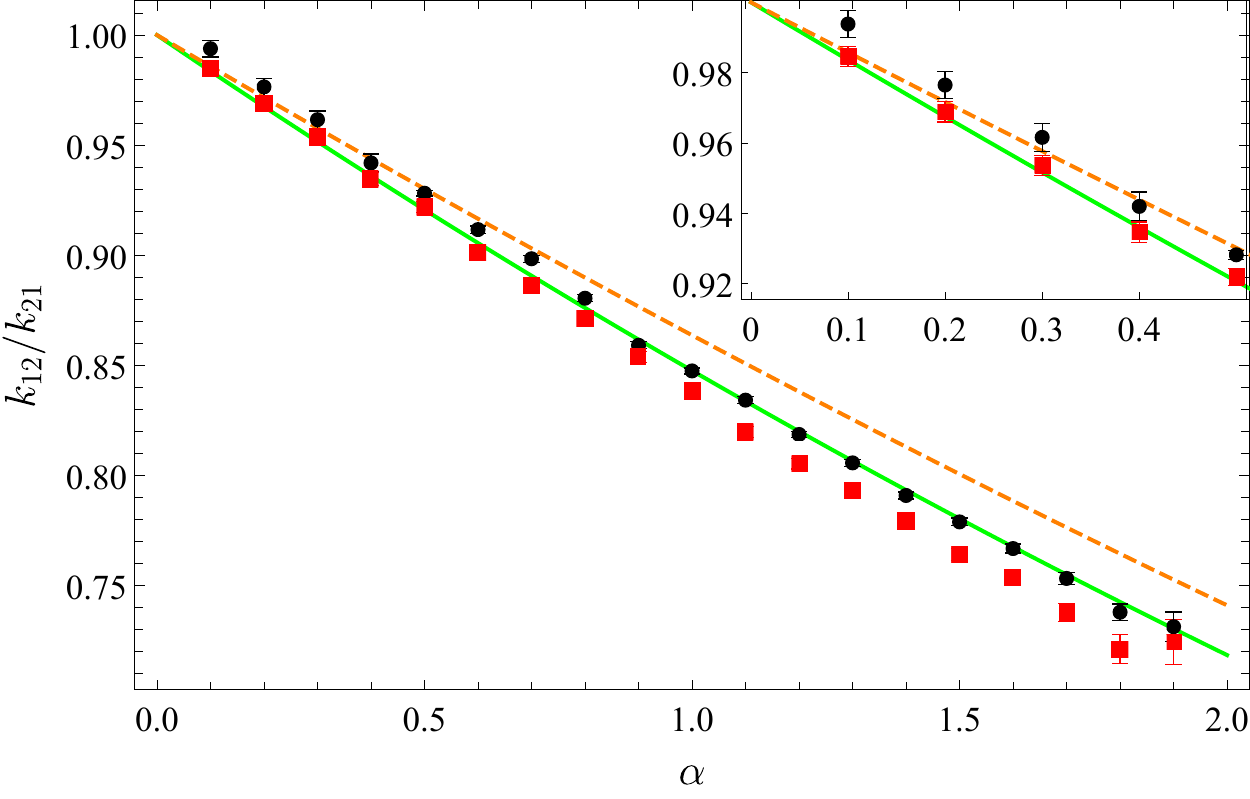}
\caption{The same as in Fig.~\ref{fig:rate}, i.e., $\kappa(\alpha)$, for various space restrictions.
Black dots (\textcolor{black}{$\bullet$}) represent unrestricted motion, while red squares (\textcolor{red}{$\blacksquare$}) correspond to the motion restricted by reflecting boundaries placed in the minima of the potential.
Solid lines present scallings given by Eq.~(\ref{eq:transitionRate}) (green solid line) and Eq.~(\ref{eq:ratioReturn}) (orange dashed line).
Simulations parameters $h_1=8$, $h_2=12$, $l_1=1$, $l_2=1$, $\gamma=1$ and $\sigma=0.2.$
}
\label{fig:rateReversal}
\end{figure}

Lack of the full agreement between predictions of Eq.~(\ref{eq:transitionRate}) and numerical results, for small $\alpha$, post the question about validity of all undertaken assumptions used to derive Eq.~(\ref{eq:transitionRate}).
First of all, the potential (\ref{eq:potential}) is not completely impenetrable at large $|x|$.
A random walker can explore outer parts of the potential corresponding to $x<-l_1$ or $x>l_2$.
\bdt{This could indicate why}
the lack of full agreement is recorded for small $\alpha$, for which the central part of the velocity distribution is narrower, and its tails are heavier.
For the large enough velocity directing outwards of the barrier top, a particle may explore the outer part of the potential and still overpass the potential barrier, see the top panel of Fig.~\ref{fig:trajectories}.
To verify \bdt{the role of exploration of outer parts of the potential}, the potential~(\ref{eq:potential}) was modified by the addition of reflecting boundaries in (i) minima of the potential, i.e., at $-l_1$ and $l_2$, or (ii) at the same distance from the potential minima as the potential barrier i.e., at $-2l_1$ and $2l_2$.
The first option improves the agreement for small $\alpha$, see red squares in Fig.~\ref{fig:rateReversal}.
At the same time, it destroys the agreement for  $\alpha \to 2$.
In the scenario (ii), ratios of transition rates are indistinguishable \bdt{(results not shown)} from results obtained in the unrestricted dynamics, see Fig.~\ref{fig:rate}.
This is in accordance with the observed dependence of $x(t)$, see Fig.~\ref{fig:trajectories}, which is restricted to $|x(t)|<2$.

Placing reflecting boundaries in minima of the potential confirms that, indeed, the differences \bdt{for small $\alpha$ between the scaling given by Eq.~(\ref{eq:transitionRate}) and numerical simulations in Fig.~\ref{fig:rate}} come from particles having the velocity directed outwards from the potential barrier.
% Large enough velocity in the outer direction is capable of producing an excursion to the outer part of the potential.
% After such an excursion the particle can return to the minimum of the potential well and cross the barrier top, see the top panel of Fig.~\ref{fig:trajectories}.
\bdt{Probability of recording an initial velocity pointing outward the potential barrier which is sufficient to induce a successful transition over the potential barrier can be calculated in the similar manner as in Eq.~(\ref{eq:transitionProbability}), but this time a particle moves along a different (longer) path.}
We can assume that the particle reverses its motion at $|x|=2l_i$, i.e., at $-2l_1$ or $2l_2$, because  as it was demonstrated in the scenario (ii) introduction of reflecting boundaries placed at $-2l_1$ and $2l_2$ produced the same results as unrestricted dynamics, see also Fig.~\ref{fig:trajectories}.
Therefore, the trajectory length is $3l_i$ and Eq.~(\ref{eq:v0}) is replaced by
\begin{equation}
    P(v>v_0) \sim \left( \sqrt{\frac{2\Delta E_i}{m}} + 3 \frac{\gamma}{m} l_i \right) ^{-\alpha}.
    \label{eq:transitionProbability3L}
\end{equation}
Finally, \bdt{taking into  account that the initial velocity can be directed towards or outwards the potential barrier}, from Eqs.~(\ref{eq:transitionProbability}) and~(\ref{eq:transitionProbability3L}), the ratio of transition rates reads
\begin{eqnarray}
\label{eq:ratioReturn}
\kappa % & = &     \frac{k_{12}}{k_{21}}   \\ \nonumber
%& = &   
= \frac{\left( \sqrt{2\Delta E_1} + \gamma l_1 \right) ^{-\alpha} +\left( \sqrt{2\Delta E_1} + 3 \gamma l_1 \right) ^{-\alpha}}{\left( \sqrt{2\Delta E_2} + \gamma l_2 \right) ^{-\alpha} +\left( \sqrt{2\Delta E_2} + 3 \gamma l_2 \right) ^{-\alpha}}.
\end{eqnarray}
Fig.~\ref{fig:rateReversal} compares scalings given by Eq.~(\ref{eq:transitionRate}) and Eq.~(\ref{eq:ratioReturn}) with results of numerical simulations for the  unrestricted  (black dots) and the restricted space (red squares), i.e., the interval $[-l_1,l_2]$.
For $\alpha>1$ agreement between simulations in the unrestricted space (black dots) and Eq.~(\ref{eq:transitionRate}) is clearly visible, as it was already \bdt{presented in Fig.~\ref{fig:rate} and  discussed within this section}.
For small $\alpha$, one might observe, that results of numerical simulations \bdt{are closer  to predictions of Eq.~(\ref{eq:ratioReturn}) than to the scaling given by Eq.~(\ref{eq:transitionRate})}, corroborating that, indeed, a part of trajectories explores outer ($|x|>|l_i|$) parts of the space.
% Exploration of outer parts of the potential is responsible for violation of Eq.~(\ref{eq:transitionRate}).
\bdt{This effect is further confirmed by the restricted motion with $\alpha<0.5$ which, up to numerical precision, follow the prediction of Eq.~(\ref{eq:transitionRate}).}
Therefore, results obtained for the dynamics in the unrestricted space (black dots) interpolates between scalings given by Eq.~(\ref{eq:ratioReturn}) (small $\alpha$, see the inset of Fig.~\ref{fig:rateReversal}) and Eq.~(\ref{eq:transitionRate}) (large $\alpha$, see the main plot in Fig.~\ref{fig:rateReversal}) with some points, corresponding to intermediate $\alpha$, laying between these two curves.
As already mentioned, results of simulations with reflecting boundaries placed in minima of the potential (red squares) follow scaling given by Eq.~(\ref{eq:transitionRate}) for small $\alpha$ only.
\bdt{Contrary to $\alpha<1$, for $\alpha>1$, the introduction of the reflecting boundaries destroys the agreement with the theoretical scaling.
The disagreement stems from two effects: (i) with increasing $\alpha$ spikes become weaker and more frequent and (ii) bounded fluctuations play a larger role.
Consequently, a particle is most likely to be found not in the potential minimum but closer to the barrier.
This in turn effectively reduces the width and the height of the potential barrier.
}

\begin{figure}[h!]
\centering
\includegraphics[width=0.95\columnwidth]{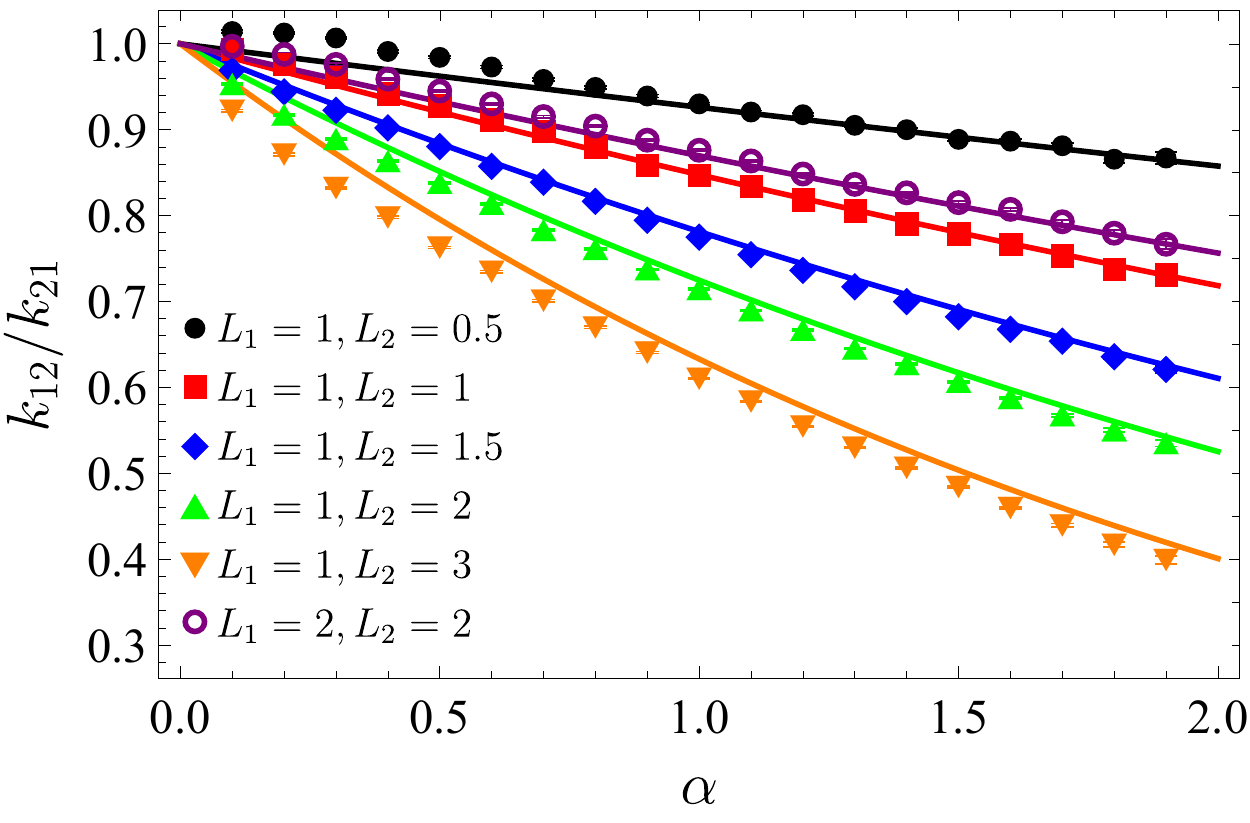}  
\caption{
The same as in Fig.~\ref{fig:rate}, i.e., $\kappa(\alpha)$, for various distances between potential minima and the maximum.
% Various points depict results of numerical simulations with various lengths  $l_1$ and $l_2$, while solid lines show the scaling given by Eq.~(\ref{eq:transitionRate}).
Simulation parameters $h_1=8$, $h_2=12$, $\gamma=1$ and $\sigma=0.2$.
}
\label{fig:rateWidth}
\end{figure}

Formula~(\ref{eq:transitionRate}) indicates that the ratio of transition rates depends both on the barrier heights and distances between minima and the maximum of the potential.
So far we have explored the \bdt{validity of Eq.~(\ref{eq:transitionRate}) for various heights of the potential barrier.
Now, we study the correctness of the scaling predicted by Eq.~(\ref{eq:transitionRate}) on changes in the distance between minima and the maximum of the potential.}
Fig.~\ref{fig:rateWidth} shows ratios of transition rates for various widths $l_1$ and $l_2$ with fixed depths $h_1=8$, $h_2=12$ and $\sigma=0.2$.
In general, results of computer simulations qualitatively follow the scaling given by Eq.~(\ref{eq:transitionRate}).
Nevertheless, quantitative deviations are especially well visible in situations when $l_1/l_2 \gg 1$, e.g., $l_1/l_2 =2$ or $l_1/l_2 =3$.

\begin{figure}[H]
\centering
\includegraphics[width=0.95\columnwidth]{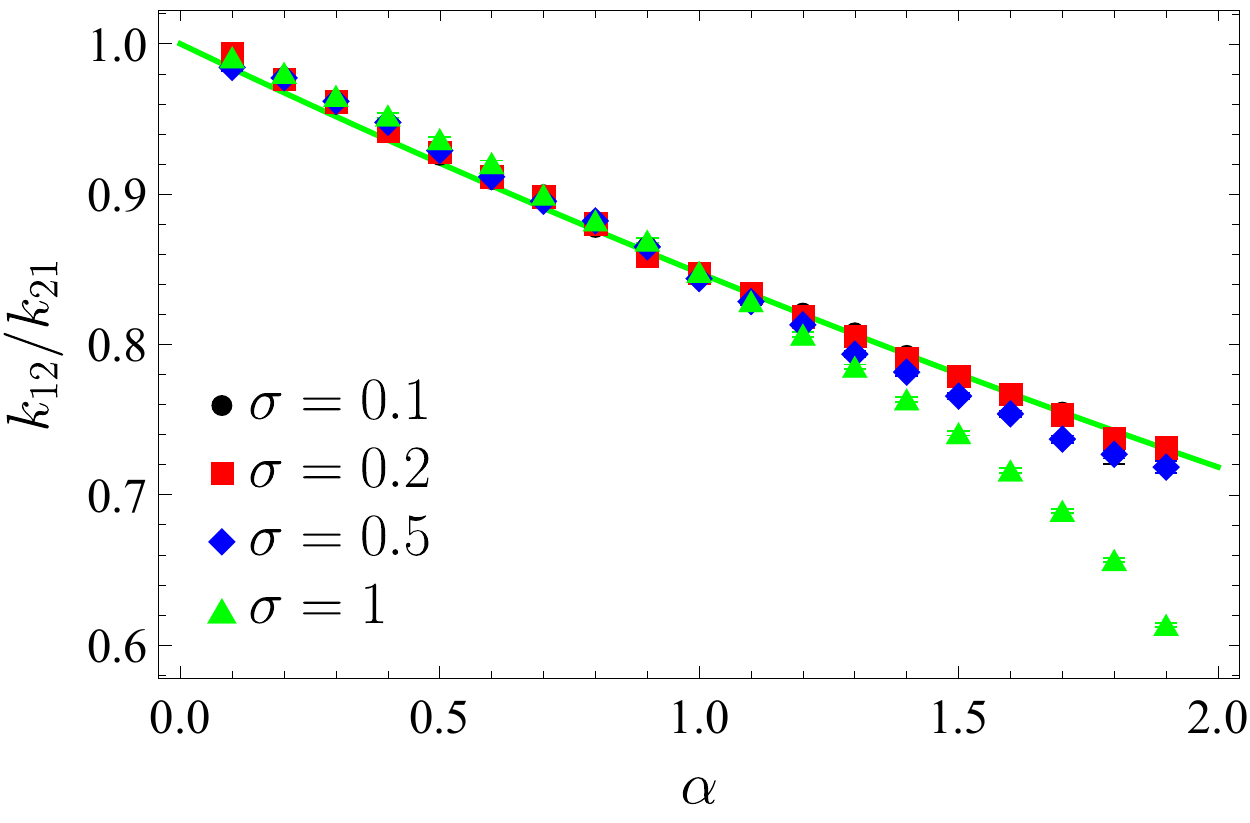}
\caption{
The same as in Fig.~\ref{fig:rate}, i.e., $\kappa(\alpha)$, for various values of the scale parameter.
% Various points depict results of numerical simulations with various scale parameters $\sigma$, while  the solid line shows the scaling given by Eq.~(\ref{eq:transitionRate}).
Simulation parameters $h_1=8$, $h_2=12$, $l_1=1$, $l_2=1$ and $\gamma=1$.
}
\label{fig:rateSigma}
\end{figure}

The ratio of transition rates, see Eq.~(\ref{eq:transitionRate}), was derived in the weak noise ($\sigma \to 0$) limit.
Nevertheless, computer simulations have confirmed the validity of Eq.~(\ref{eq:transitionRate}) for small but finite values of the scale parameter $\sigma$.
Therefore, we have checked if results obtained under the weak noise approximation holds for larger $\sigma$, and how the ratio of transition rates behaves in this case.
Fig.~\ref{fig:rateSigma} presents ratios of transition rates for various values of the scale parameter $\sigma$.
For $\alpha < 1$ results for all used values of $\sigma$ \bdt{follow the scaling given by Eq.~(\ref{eq:transitionRate})}.
\bdt{The situation changes for $\alpha>1$}, because the agreement between results of computer simulations and Eq.~(\ref{eq:transitionRate}) is recorded only for small values of $\sigma$, e.g., $\sigma=0.1$ and $\sigma=0.2$.
Results for $\sigma=0.5$ are still very close to the scaling given by Eq.~(\ref{eq:transitionRate}), however, one may observe that the ratio of transition rates is slightly smaller than the weak noise prediction.
This deviation amplifies with the increasing $\sigma$, and for $\sigma=1$ results diverge quickly from the weak noise scaling.
The amplification of deviations is very similar to the behavior in the overdamped regime \cite{capala2020athermal} and can be attributed to the violation of the weak noise approximation, i.e., for large $\sigma$, transitions occur not only via a single change in the velocity, but also due to a series of smaller ``kicks''.
\bdt{Consequently, Eq.~(\ref{eq:transitionProbability}) cannot be straight forward applied.}

\begin{figure}[h!]
\centering
\includegraphics[width=0.99\columnwidth]{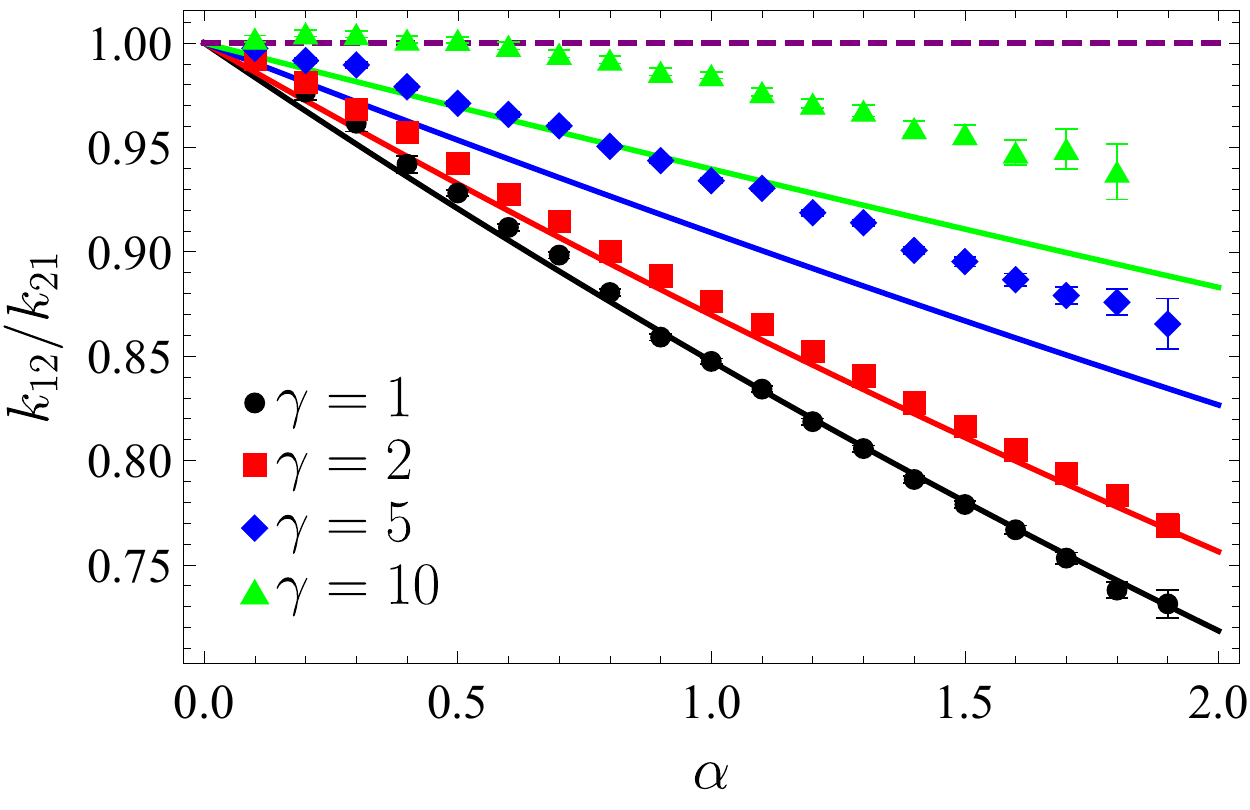}
\caption{
The same as in Fig.~\ref{fig:rate}, i.e., $\kappa(\alpha)$, for various values of the friction coefficient $\gamma$.
% Various points depict results of numerical simulations with various values of damping  $\gamma$, while  solid lines show the scaling given by Eq.~(\ref{eq:transitionRate}).
Simulation parameters $h_1=8$, $h_2=12$, $l_1=1$, $l_2=1$ and $\sigma=0.2$.
The purple dashed line corresponds to overdamped scaling given by Eq.~(\ref{eq:overdampedRatio}).
}
\label{fig:rateGamma}
\end{figure}

Finally, we \bdt{estimate numerically} the ratio of transition rates for the increasing damping strength.
In the limit of $\gamma \to \infty$, Eq.~(\ref{eq:fullLangevin}) correctly reduces to the overdamped Langevin equation, for which the ratio of transition rates is given by Eq.~(\ref{eq:overdampedRatio}).
Therefore, from Eq.~(\ref{eq:transitionRate}) one might expect a smooth, steady transition to the ratio of transition rates for the overdamped limit, i.e., to Eq.~(\ref{eq:overdampedRatio}).
As it is clearly visible from Fig.~\ref{fig:rateGamma}, the transition is not smooth.
With the increasing $\gamma$, the ratio of transition rates increases.
For small values of the friction parameter $\gamma$, simulation results reproduce predictions of Eq.~(\ref{eq:transitionRate}), but with the increasing $\gamma$ results of simulations deviate from the prediction given by Eq.~(\ref{eq:transitionRate}).
In particular, for $\gamma=5$, numerically estimated ratios of transition rates follow predictions of Eq.~(\ref{eq:transitionRate}) with $\gamma=10$ almost precisely.
For $\gamma=10$, with $\alpha<0.75$, the ratio of transition rates reached the overdamped limit.
Simultaneously,  for $\alpha>0.75$, $\kappa(\alpha)$ significantly deviates both from the underdamped and overdamped scalings.
In overall, this indicates that the overdamped limit is reached already for a finite damping, but the critical value of $\gamma$ depends on the stability index $\alpha$.
\bdt{In particular}, for small $\alpha$ the overdamped limit is reached faster.
\bdt{Otherwise for $\gamma$ smaller than critical, results are sensitive not only to the stability index $\alpha$ but also to the damping strengths, see Fig.~\ref{fig:rateGamma}.}

%%%%%%%%%%%%%%%%%%%%%%%%%%%%%%%%%%%%%%%%%%%%%%%%%%%%%%%%%%%%%%%%%%%%%%%%%%%%%%%%%%%%%%
%%%%%%%%%%%%%%%%%%%%%%%%%%%%%%%%%%%%%%%%%%%%%%%%%%%%%%%%%%%%%%%%%%%%%%%%%%%%%%%%%%%%%%

% \clearpage
\section{Summary and Conclusions\label{sec:summary}}

The  escape of a particle from the potential well is possible due to action of the noise.
The escape protocol is sensitive both to the noise type (Gaussian versus L\'evy) and dynamic type (overdamped versus underdamped).
In the overdamped regime a particle is fully characterized by the position.
The particle can jump over the potential barrier or surmount it.
Therefore, during the escape from the potential well a particle is either waiting for the strong enough noise pulse (L\'evy) or for a sequence of small kicks (Gaussian driving).
In the underdamped regime, the particle needs to harvest energy which is sufficient to overpass the potential barrier.
Analogously like in the underdamped regime, the particle steadily accumulates energy (Gaussian noise) or it waits for the abrupt jump in the velocity (L\'evy driving).

The most significant difference between L\'evy noise and Gaussian noise induced escape is recorded in the overdamped case.
L\'evy process with $\alpha<2$ has discontinuous trajectories, while paths of the Brownian motion are continuous.
In the weak noise limit, under $\alpha$-stable noise, the ratio of reaction rates depends on the barrier widths, because the particle waits for the jump which is long enough \bdt{as it is the main escape protocol}.
Consequently, the escape time is insensitive to the barrier height.
The escape under Gaussian white noise follows a completely different scenario.
The particle escapes via a sequence of short jumps, therefore the transition rate is sensitive to the barrier height.

The underdamped regime is very different from the overdamped regime, because in the underdamped regime the trajectory $x(t)$ is continuous both under L\'evy and Gaussian drivings.
The escaping particle needs to harvest sufficient energy to pass over the potential barrier.
Therefore, the ratio of the escape rates is sensitive to the barrier height, also in the weak noise limit, both under Gaussian and L\'evy drivings, as the barrier height defines the amount of energy which needs to be accumulated.
\bdt{Various regimes (overdamped and underdamped) and various drivings (Gaussian and L\'evy) are compared in Tab.~\ref{tab:ratio}.}

\begin{table}[!h]
    \centering
    \begin{tabularx}{0.95\columnwidth}{
   >{\centering\arraybackslash}X
  || >{\centering\arraybackslash}X
  | >{\centering\arraybackslash}X  }
         & Gaussian  & L\'evy  \\
         \hline \hline
       overdamped  & ratio of transition rates depends on difference of the potential well depths  & ratio of transition rates depends on ratio of the potential well widths\\
       \hline
       underdamped  & ratio of transition rates depends on the potential barrier heights & ratio of transition rates depends on the potential barriers heights and widths\\
    \end{tabularx}
    \caption{The compilation of information on dependence of the ratio of transition rates in double-well potentials for various escape scenarios (Gaussian driving vs L\'evy driving) and various regimes (overdamped vs underdamped).}
    \label{tab:ratio}
\end{table}

In the weak noise limit, under action of L\'evy noise a particle typically escapes due to a single rapid change in the velocity.
Using asymptotic properties of $\alpha$-stable densities, we have derived the formula for the ratio of escape rates, \bdt{see  Eq.~(\ref{eq:transitionRate}), which is the main result of current research.}
\bdt{It shows that} the ratio of the escape rates depends both on the barrier widths and heights, but the sensitivity to the barrier width is larger.
In the limit of the large friction the derived formula correctly reduces to the result already known for the overdamped dynamics, i.e., the ratio of transition rates depends on the width of the potential barrier only \cite{ditlevsen1999}.
% \karol{Eeee... Wynik Biera jest prawdziwy tylko dla procesu dyskretnego w czasie. Limit przetłumiony naszego problemu to tylko Ditlevsen. Poza tym w granicy silnego tłumienia nie otrzymujemy wyniku Biera, bo dla nas energia znika, a dla niego eksploduje. Osobiście zacytowałbym raczej \cite{ditlevsen1999}}
The obtained formula works very well under the assumption that the studied process, more precisely its spatial part, can be approximated as the two state process.
Consequently, the potential barrier separating minima and outer parts of the potential need to be steep enough.
Deviations from the derived formula are especially visible when a particle position is not restricted to the vicinity of the potential minima.
It happens when the restoring force is not large enough, or noise cannot be considered as weak.

%%%%%%%%%%%%%%%%%%%%%%%%%%%%%%%%%%%%%%%%%%%%%%%%%%%%%%%%%%%%%%%%%%%%%%%%%%%%%%%%%%%%%%
%%%%%%%%%%%%%%%%%%%%%%%%%%%%%%%%%%%%%%%%%%%%%%%%%%%%%%%%%%%%%%%%%%%%%%%%%%%%%%%%%%%%%%

\section*{Acknowledgements}

This research was supported in part by PLGrid Infrastructure and by the National Science Center (Poland) grant 2018/31/N/ST2/00598.

% This project was supported by the National Science Center (Poland) grant 2018/31/N/ST2/00598.
% This research was supported in part by PLGrid Infrastructure.

%%%%%%%%%%%%%%%%%%%%%%%%%%%
\appendix
% \section{Weak noise approximation}

% In order to

% The overdamped Langevin equation
% \begin{equation}
%     \dot{x}=-V'(x)+\zeta(t)
% \end{equation}
% can be rewritten in the following more formal form
% \begin{eqnarray}
% x(t)=x(0)-\int_0^t V'(x(s)) ds + \varepsilon L(t)
% \end{eqnarray}

% The $\alpha$-stable process, see Eq.~(\ref{eq:fcharakt}), is the special type of the L\'evy process.
% The L\'evy process is a stochastic process process with independent, stationary increments.
% Moreover, increments in time intervals of the same length are independent, identically distributed.
% The $\alpha$-stable white noise is the formal time derivative of the $\alpha$-stable process $L(t)$.

% According to the L\'evy-Khintchine formula the characteristic function of the L\'evy process is given by
% \begin{widetext}
% \begin{equation}
%     \langle \exp[ikL(t)] \rangle = \exp\left\{ -t \left[   iak  -d \frac{k^2}{2}  + \int_{\mathbb{R} \setminus \{0\} }  \left( \mbox{e}^{iky} - 1 - i k y  \mathbbm{1}(|y|-1)  \right) \frac{dy}{|y|^{1+\alpha}}   \right]  \right\},
% \end{equation}
% \end{widetext}
% where:
% $\mathbbm{1}$ is the indicator function,
% $a$ ($a>0$) -- constant drift, $d$ ($d>0$) -- variance.
% From L\'evy-Khintchine formula it implies that L\'evy process is a sum of independent process: a process with the linear drift $a$,
% a standard Brownian motion with variance $d$, and a $\alpha$-stable motion with $0<\alpha<2$.

%%%%%%%%%%%%%%%%%%%%%%%%%%%
\section{Velocity distribution \label{sec:velocity}}

\bdt{
In the regime of full dynamics, under linear friction, the velocity evolves according to
\begin{equation}
    \frac{dv}{dt}= -\gamma v -V'(x) + \sigma \zeta(t),
    \label{eq:velocity}
\end{equation}
see Eq.~(\ref{eq:fullLangevin}).
If we omit the deterministic force $-V'(x)$ in Eq.~(\ref{eq:velocity}), the Langevin equation is associated with the following velocity-fractional Smoluchowski-Fokker-Planck equation
\begin{equation}
    \frac{\partial P (v,t) }{\partial t} = \frac{\partial  }{\partial v} \left[ \gamma v P(v,t)  \right] + \sigma^\alpha \frac{\partial^\alpha P (v,t) }{\partial |v|^\alpha}.
\end{equation}
In the stationary state one has
\begin{equation}
    0 = \frac{d  }{d v} \left[ \gamma v P(v)  \right] + \sigma^\alpha \frac{d^\alpha P (v) }{d |v|^\alpha}.
    \label{eq:stationary}
\end{equation}
In the Fourier space Eq.~(\ref{eq:stationary}) reads
\begin{equation}
    \gamma k \frac{d \hat{P}(k) }{d k } = -\sigma^\alpha |k|^\alpha \hat{P}(k),
\end{equation}
where $\hat{P}(k)$ is the Fourier transform $\hat{P}(k)=\int_{-\infty}^\infty P(v) e^{ikv} dv$.
The characteristic function $\hat{P}(k)$ of the stationary distribution $P(v)$ satisfies
\begin{equation}
    \frac{d \hat{P}(k) }{d k} = -\frac{\sigma^\alpha}{\gamma} \sign( k) |k|^{\alpha-1} \hat{P}(k).
    \label{eq:st-eq}
\end{equation}
The solution of Eq.~(\ref{eq:st-eq}) is given by
\begin{equation}
    \hat{P}(k) = \exp\left[ -\frac{\sigma^\alpha }{\gamma\alpha}   |k|^\alpha \right],
\end{equation}
which is the characteristic function of the symmetric $\alpha$-stable distribution, see Eq.~(\ref{eq:fcharakt}), with the scale parameter $\sigma'$
\begin{equation}
\sigma' = \frac{\sigma}{(\gamma \alpha)^{1/\alpha}}    .
\label{eq:modsigma}
\end{equation}
With the increasing $\gamma$, the stationary distribution becomes narrower.
For instance, for the Cauchy noise ($\alpha=1$), the stationary density is the Cauchy distribution
\begin{equation}
    P(v) = \frac{1}{\pi} \frac{\sigma'}{(\sigma')^2+v^2}.
    \label{eq:modpv}
\end{equation}
In more general cases, the asymptotic behavior of $P(v)$ is given by
\begin{eqnarray}
P(v) \sim  \sigma^\alpha  \frac{\Gamma(\alpha+1)}{\pi} \sin \frac{\pi\alpha}{2} \times \frac{1}{|v|^{\alpha+1}}.
\label{eq:asymptoti}
\end{eqnarray}
Eq.~(\ref{eq:tails}) implies from Eq.~(\ref{eq:asymptoti}).
}

%\bibliographystyle{prsty}
%\bibliography{core-bibliography}

\def\url#1{}

\end{document}